# Designing and Mining Multi-Terabyte Astronomy Archives: The Sloan Digital Sky Survey

Alexander S. Szalay
Peter Kunszt
Ani Thakar
Dept. of Physics and Astronomy, The Johns Hopkins University, Baltimore, MD 21218

Jim Gray
Microsoft Research, San Francisco, CA 94105

June 1999

Technical Report
MS-TR-99-30

# Designing and Mining Multi-Terabyte Astronomy Archives: The Sloan Digital Sky Survey

Alexander S. Szalay, szalay@jhu.edu
Peter Kunszt, kunszt@pha.jhu.edu
Ani Thakar, thakar@pha.jhu.edu
Dept. of Physics and Astronomy, The Johns Hopkins University, Baltimore, MD 21218
Jim Gray, gray@Microsoft.com
Microsoft Research, San Francisco, CA 94105

## Abstract

The next-generation astronomy digital archives will cover most of the universe at fine resolution in many wavelengths, from X-rays to ultraviolet, optical, and infrared. The archives will be stored at diverse geographical locations. One of the first of these projects, the Sloan Digital Sky Survey (SDSS) will create a 5-wavelength catalog over 10,000 square degrees of the sky (see http://www.sdss.org/). The 200 million objects in the multi-terabyte database will have mostly numerical attributes, defining a space of 100+ dimensions. Points in this space have highly correlated distributions.

The archive will enable astronomers to explore the data interactively. Data access will be aided by a multidimensional spatial index and other indices. The data will be partitioned in many ways. Small *tag* objects consisting of the most popular attributes speed up frequent searches. Splitting the data among multiple servers enables parallel, scalable I/O and applies parallel processing to the data. Hashing techniques allow efficient clustering and pairwise comparison algorithms that parallelize nicely. Randomly sampled subsets allow debugging otherwise large queries at the desktop. Central servers will operate a data pump that supports sweeping searches that touch most of the data. The anticipated queries require special operators related to angular distances and complex similarity tests of object properties, like shapes, colors, velocity vectors, or temporal behaviors. These issues pose interesting data management challenges.

## Introduction

Astronomy is about to undergo a major paradigm shift. Data gathering technology is riding Moore's law: volumes are doubling quickly. Data sets are becoming larger, and more homogeneous. For the first time data acquisition and archival is being designed for online interactive analysis. In a few years it will be much easier to download a detailed sky map or object class catalog, than wait several months to access a telescope that is often quite small.

Several multi-wavelength projects are under way: SDSS, GALEX, 2MASS, GSC-2, POSS2, ROSAT, FIRST and DENIS. Each is surveying a large fraction of the sky. Together they will yield a *Digital Sky*, of interoperating multi-terabyte databases. In time, more catalogs will be added and linked to the existing ones. Query engines will become more sophisticated, providing a uniform interface to all these datasets. In this era, astronomers will have to be just as familiar with mining data as with observing on telescopes.

## The Sloan Digital Sky Survey

The Sloan Digital Sky Survey (SDSS) will digitally map about half of the Northern sky in five spectral bands from ultraviolet to the near infrared. It is expected to detect over 200 million objects. Simultaneously, it will measure redshifts for the brightest million galaxies (see http://www.sdss.org/).

The SDSS is the successor to the Palomar Observatory Sky Survey (POSS), which provided a standard reference data set to all of astronomy for the last 40 years. Subsequent archives will augment the SDSS and will interoperate with it. The SDSS project thus consists of not only of building the hardware, and reducing and calibrating the data, but also includes software to classify, index, and archive the data so that many scientists can use it.

The SDSS will revolutionize astronomy, increasing the amount of information available to researchers by several orders of magnitude. The SDSS archive will be large and complex: including textual information, derived parameters, multi-band images, spectra, and temporal data. The catalog will allow astronomers to study the evolution of the universe in great detail. It is intended to serve as the standard reference for the next several decades. After only a month of operation, SDSS found the two most distant known quasars. With more data, other exotic properties will be easy to mine from the datasets.

The potential scientific impact of the survey is stunning. To realize this potential, data must be turned into knowledge. This is not easy – the information content of the survey will be larger than the entire text contained in the Library of Congress.

The SDSS is a collaboration between the University of Chicago, Princeton University, the Johns Hopkins University, the University of Washington, Fermi National Accelerator Laboratory, the Japanese Participation Group, the United States Naval Observatory, and the Institute for Advanced Study, Princeton, with additional funding provided by the Alfred P. Sloan Foundation, NSF and NASA. The SDSS project is a collaboration between scientists working in diverse areas of astronomy, physics and computer science. The survey will be carried out with a suite of tools developed and built especially for this project – telescopes, cameras, fiber spectrographic systems, and computer software.

SDSS constructed a dedicated 2.5-meter telescope at Apache Point, New Mexico, USA. The telescope has a large, flat focal plane that provides a 3-degree field of view. This design balances the areal coverage of the instrument against the detector's pixel resolution.

The survey has two main components: a photometric survey, and a spectroscopic survey. The photometric survey is produced by drift scan imaging of 10,000 square degrees centered on the North Galactic Cap using five broad-band filters that range from the ultra-violet to the infra-red. The effective exposure is 55 sec. The photometric imaging uses an array of 30x2Kx2K Imaging CCDs, 22 Astrometric CCDs, and 2 Focus CCDs. Its 0.4 arcsec pixel size provides a full sampling of the sky. The data rate from the 120 million pixels of this camera is 8 Megabytes per second. The cameras can only be used under ideal conditions, but during the 5 years of the survey SDSS will collect more than 40 Terabytes of image data.

The spectroscopic survey will target over a million objects chosen from the photometric survey in an attempt to produce a statistically uniform sample. Due to the expansion of the universe, the Doppler-shift in the spectral lines of the galaxies is a direct measure of their distance. The result of the spectroscopic survey will be a three-dimensional map of the galaxy distribution, in a volume several orders of magnitude larger than earlier maps.

The primary targets will be galaxies, selected by a magnitude and surface brightness limit in the *r* band. This sample of 900,000 galaxies will be complemented with 100,000 very red galaxies, selected to include the brightest galaxies at the cores of clusters. An automated algorithm will select 100,000 quasar candidates for spectroscopic follow-up, creating the largest uniform quasar survey to date. Selected objects from other catalogs (FIRST, ROSAT) will also be targeted.

The spectroscopic observations will be done in overlapping 3? circular "tiles". The tile centers are determined by an optimization algorithm, which maximizes overlaps at areas of highest target density. The spectroscopic survey will utilize two multi-fiber medium resolution spectrographs, with a total of 640 optical fibers. Each fiber is 3 seconds of arc in diameter, that provide spectral coverage from 3900 - 9200 Å. The system can measure 5000 galaxy spectra per night. The total number of galaxy spectra known to astronomers today is about 100,000 – only 20 nights of SDSS data! Whenever the Northern Galactic cap is not accessible, SDSS repeatedly images several areas in the Southern Galactic cap to study fainter objects and identify variable sources.

SDSS has also been developing the software necessary to process and analyze the data. With construction of both hardware and software largely finished, the project has now entered a year of integration and testing. The survey itself will take 5 to 7 years to complete, depending on weather.

### *The Data Products*

The SDSS will create four main data sets: (1) a photometric catalog, (2) a spectroscopic catalog, (3) images, and (4) spectra. The photometric catalog is expected to contain about 500 distinct attributes for each of one hundred million galaxies, one hundred million stars, and one million quasars. These include positions, fluxes, radial profiles, their errors, and information related to the observations. Each object will have an associated image cutout ("atlas image") for each of the five filters. The spectroscopic catalog will contain identified emission and absorption lines, and one-dimensional spectra for 1 million galaxies, 100,000 stars, and 100,000 quasars. Derived custom catalogs may be included, such as a photometric cluster catalog, or quasar absorption line catalog. In addition there will be a compressed 1TB Sky Map. As shown in Table 1, these products are about 3TB.

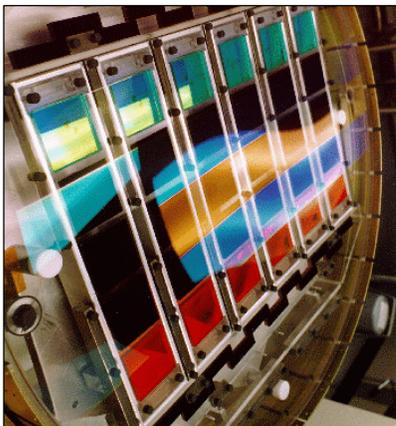

Figure 1: The SDSS photometric camera with the 5x6 CCD array contains 120 million pixels. The CCDs in each row have a different filter attached.

| Table 1. Sizes of various SDSS datasets | | |
|---|---|---|
| *Product* | *Items* | *Size* |
| Raw observational data | - | 40 TB |
| Redshift Catalog | $10^6$ | 2 GB |
| Survey Description | $10^5$ | 1 GB |
| Simplified Catalog | $3 \times 10^8$ | 60 GB |
| 1D Spectra | $10^6$ | 60 GB |
| Atlas Images | $10^9$ | 1.5 TB |
| Compressed Sky Map | $5 \times 10^5$ | 1.0 TB |
| Full photometric catalog | $3 \times 10^8$ | 400 GB |

The collaboration will release this data to the public after a period of thorough verification. This public archive is expected to remain the standard reference catalog for the next several decades. This long-lifetime presents design and legacy problems. The design of the SDSS archival system must allow the archive to grow beyond the actual completion of the survey. As *the* reference astronomical data set, each subsequent astronomical survey will want to cross-identify its objects with the SDSS catalog, requiring that the archive, or at least a part of it, be dynamic with a carefully defined schema and metadata.

### *The SDSS Archives*

Observational data from the telescopes is shipped on tapes to Fermi National Laboratory (FNAL) where it is reduced and stored in the *Operational Archive (OA)*, protected by a firewall, accessible only to personnel working on the data processing.

Data in the operational archive is reduced and calibrated via method functions. Within two weeks the calibrated data is published to the *Science Archive (SA)*. The Science Archive contains calibrated data organized for efficient science use. The SA provides a custom query engine that uses multidimensional indices. Given the amount of data, most queries will be I/O limited, thus the SA design is based on a scalable architecture, ready to use large numbers of cheap commodity servers, running in parallel.

Science archive data is replicated to Local Archives (LA) within another two weeks. The data gets into the public archives (MPA, PA) after approximately 1-2 years of science verification, and recalibration (if necessary).

A WWW server will provide public access. The public will be able to see project status and see various images including the 'Image of the week'.

The Science Archive and public archives employ a three-tiered architecture: the user interface, an intelligent query engine, and the data warehouse. This distributed approach provides maximum flexibility, while maintaining portability, by isolating hardware specific features. Both the Science Archive and the Operational Archive are built on top of Objectivity/DB, a commercial OODBMS.

Querying these archives requires a parallel and distribute query system. We have implemented a prototype query system. Each query received from the User Interface is parsed into a Query Execution Tree (QET) that is then executed by the Query Engine. Each node of the QET is either a query or a set-operation node, and returns a bag of object-pointers upon execution. The multi-threaded Query Engine executes in parallel at all the nodes at a given level of the QET.

Results from child nodes are passed up the tree as soon as they are generated. In the case of aggregation, sort, intersection and difference nodes, at least one of the child nodes must be complete before results can be sent further up the tree. In addition to speeding up the query processing, this ASAP data push strategy ensures that even in the case of a query that takes a very long time to complete, the user starts seeing results almost immediately, or at least as soon as the first selected object percolates up the tree.

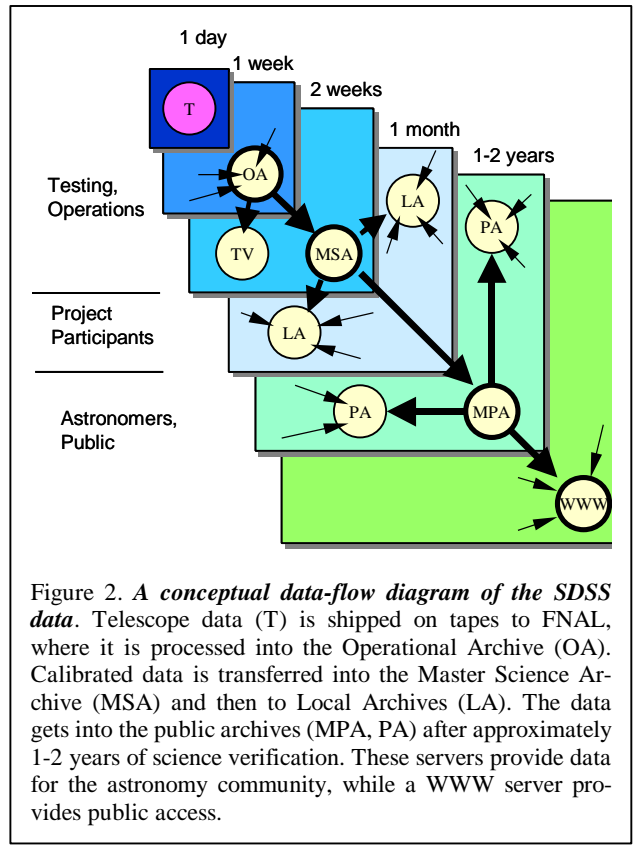

Figure 2. *A conceptual data-flow diagram of the SDSS data*. Telescope data (T) is shipped on tapes to FNAL, where it is processed into the Operational Archive (OA). Calibrated data is transferred into the Master Science Archive (MSA) and then to Local Archives (LA). The data gets into the public archives (MPA, PA) after approximately 1-2 years of science verification. These servers provide data for the astronomy community, while a WWW server provides public access.

### *Typical Queries*

The astronomy community will be the primary SDSS user. They will need specialized services. At the simplest level these include the on-demand creation of (color) finding charts, with position information. These searches can be fairly complex queries on position, colors, and other parts of the attribute space.

As astronomers learn more about the detailed properties of the stars and galaxies in the SDSS archive, we expect they will define more sophisticated classifications. Interesting objects with unique properties will be found in one area of the sky. They will want to generalize these properties, and search the entire sky for similar objects.

A common query will be to distinguish between rare and typical objects. Other types of queries will be non-local, like "*find all the quasars brighter than r=22, which have a faint blue galaxy within 5 arcsec on the sky*". Yet another type of a query is a search for gravitational lenses: *"find objects within 10 arcsec of each other which have identical colors, but may have a different brightness"*. This latter query is a typical high-dimensional query, since it involves a metric distance not only on the sky, but also in color space.

Special operators are required to perform these queries efficiently. Preprocessing, like creating regions of attraction is not practical, given the number of objects, and that the sets of objects these operators work on are dynamically created by other predicates.

### *Geometric Data Organization*

Given the huge data sets, the traditional Fortran access to flat files is not a feasible approach for SDSS. Rather non-procedural query languages, query optimizers, database execution engines, and database indexing schemes must replace traditional "flat" file processing.

This "database approach" is mandated both by computer efficiency, and by the desire to give astronomers better analysis tools.

The data organization must support concurrent complex queries. Moreover, the organization must efficiently use processing, memory, and bandwidth. It must also support the addition of new data to the SDSS as a background task that does not disrupt online access.

It would be wonderful if we could use an off-the-shelf SQL, OR, or OO database system for our tasks, but we are not optimistic that this will work. As explained presently, we believe that SDSS requires novel spatial indices and novel operators. It also requires a dataflow architecture that executes queries concurrently using multiple disks and processors. As we understand it, current systems provide few of these features. But, it is quite possible that by the end of the survey, some commercial system will provide these features. We hope to work with DBMS vendors towards this end.

### *Spatial Data Structures*

The large-scale astronomy data sets consist primarily of vectors of numeric data fields, maps, time-series sensor logs and images: the vast majority of the data is essentially geometric. The success of the archive depends on capturing the spatial nature of this large-scale scientific data.

The SDSS data has high dimensionality -- each item has thousands of attributes. Categorizing objects involves defining complex domains (classifications) in this N-dimensional space, corresponding to decision surfaces.

The SDSS teams are investigating algorithms and data structures to quickly compute spatial relations, such as finding nearest neighbors, or other objects satisfying a given criterion within a metric distance. The answer set cardinality can be so large that intermediate files simply cannot be created. The only way to analyze such data sets is to pipeline the answers directly into analysis tools. This *data flow* analysis has worked well for parallel relational database systems [DeWitt92]. We expect these data river ideas will link the archive directly to the analysis and visualization tools.

The typical search of these multi-Terabyte archives evaluates a complex predicate in k-dimensional space, with the added difficulty that constraints are not necessarily parallel to the axes. This means that the traditional indexing techniques, well established with relational databases, will not work, since one cannot build an index on all conceivable linear combinations of attributes. On the other hand, one can use the fact that the data are geometric and every object is a point in this k-dimensional space [Samet 1990a,b]. Data can be quantized into containers. Each container has objects of similar properties, e.g. colors, from the same region of the sky. If the containers are stored as clusters, data locality will be very high - if an object satisfies a query, it is likely that some of the object's "friends" will as well. There are non-trivial aspects of how to subdivide, when the data has large density contrasts [Csabai96].

These containers represent a coarse-grained density map of the data. They define the base of an index tree that tells us whether containers are fully inside, outside or bisected by our query. Only the bisected container category is searched, as the other two are wholly accepted or rejected. A prediction of the output data volume and search time can be computed from the intersection volume.

### *Indexing the Sky*

The SDSS data is too large to fit on one disk or even one server. The base-data objects will be spatially partitioned among the servers. As new servers are added, the data will repartition.

Some of the high-traffic data will be replicated among servers. It is up to the database software to manage this partitioning and replication. In the near term, designers will specify the partitioning and index schemes, but we hope that in the long term, the DBMS will automate this design task as access patterns change.

There is great interest in a common reference frame the sky that can be universally used by different astronomical databases. The need for such a system is indicated by the widespread use of the ancient constellations – the first spatial index of the celestial sphere. The existence of such an index, in a more computer friendly form will ease cross-referencing among catalogs.

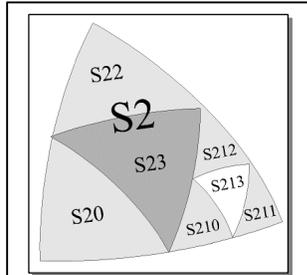

Figure 3. The hierarchical subdivision of spherical triangles, represented as a quad tree. The tree starts out from the triangles defined by an octahedron.

A common scheme, that provides a balanced partitioning for all catalogs, may seem to be impossible; but, there is an elegant solution, a 'shoe that fits all': that subdivides the sky in a hierarchical fashion.

Instead of taking a fixed subdivision, we specify an increasingly finer hierarchy, where each level is fully contained within the previous one. Starting with an octahedron *base set,* each spherical triangle can be recursively divided into 4 sub-triangles of approximately equal areas. Each sub-area can be divided further into additional four sub-areas, ad infinitum. Such hierarchical subdivisions can be very efficiently represented in the form of quad-trees. Areas in different catalogs map either directly onto one another, or one is fully contained by another (see Figure 3.)

We store the angular coordinates in a Cartesian form, i.e. as a triplet of *x,y,z* values per object. The *x,y,z* numbers represent only the position of objects on the sky, corresponding to the normal vector pointing to the object. (We can guess the distance for only a tiny fraction of the 200 million objects in the catalog.) While at first this may seem to increase the required storage (three numbers per object vs two angles,) it makes querying the database for objects within certain areas of the celestial sphere, or involving different coordinate systems considerably more efficient. This technique was used successfully by the GSC project [GSC2]. The coordinates in the different celestial coordinate systems (Equatorial, Galactic, Supergalactic, etc) can be constructed from the Cartesian coordinates on the fly.

Due to the three-dimensional Cartesian representation of the angular coordinates, queries to find objects within a certain spherical distance from a given point, or combination of constraints in arbitrary spherical coordinate systems become particularly simple. They correspond to testing linear combinations of the three Cartesian coordinates instead of complicated trigonometric expressions.

The two ideas, partitioning and Cartesian coordinates merge into a highly efficient storage, retrieval and indexing scheme. We have created a recursive algorithm that can determine which parts of the sky are relevant for a particular query [Szalay99]. Each query can be represented as a set of half-space constraints, connected by Boolean operators, all in three-dimensional space.

The task of finding objects that satisfy a given query can be performed recursively as follows. Run a test between the query polyhedron and the spherical triangles corresponding to the tree root nodes. The intersection algorithm is very efficient because it is easy to test spherical triangle intersection. Classify nodes, as fully outside the query, fully inside the query or partially intersecting the query polyhedron. If a node is rejected, that node's children can be ignored. Only the children of bisected triangles need be further investigated. The intersection test is executed recursively on these nodes. The SDSS Science Archive implemented this algorithm in its query engine [Szalay97].

### *Broader Metadata Issues*

There are several issues related to metadata for astronomy datasets. One is the database schema within the data warehouse, another is the description of the data extracted from the archive and the third is a standard representation to allow queries and data to be interchanged among several archives.

The SDSS project uses Platinum Technology's Paradigm Plus, a commercially available UML tool, to develop and maintain the database schema. The schema is defined in a high level format, and an automated script generator creates the .h files for the C++ classes, and the .ddl files for Objectivity/DB. This approach enables us to easily create new data model representations in the future (SQL, IDL, XML, etc).

About 20 years ago, astronomers agreed on exchanging most of their data in self-descriptive data format. This format, FITS, standing for the Flexible Image Transport

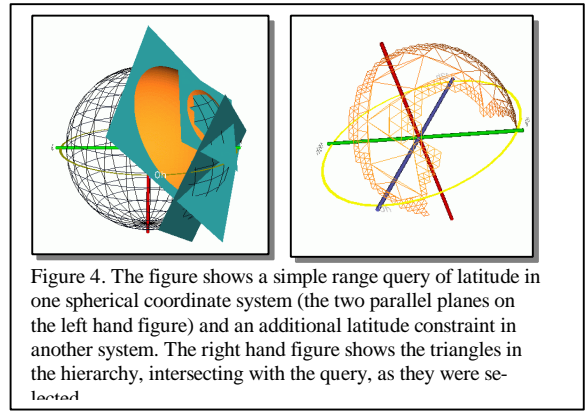

Figure 4. The figure shows a simple range query of latitude in one spherical coordinate system (the two parallel planes on the left hand figure) and an additional latitude constraint in another system. The right hand figure shows the triangles in the hierarchy, intersecting with the query, as they were selected.

System [Wells81] was primarily designed to handle images. Over the years, various extensions supported more complex data types, both in ASCII and binary form. FITS format is well supported by all astronomical software systems. The SDSS pipelines exchange most of their data as binary FITS files.

Unfortunately, FITS files do not support streaming data, although data could be blocked into separate FITS packets. We are currently implementing both an ASCII and a binary FITS output stream, using such a blocked approach.

We expect large archives to communicate with one another via a standard, easily parseable interchange format. We plan to define the interchange formats in XML, XSL, and XQL.

### Data Loading

The Operational Archive exports calibrated data to the Science Archive as soon as possible. Datasets are sent in coherent chunks. A chunk consists of several segments of the sky that were scanned in a single night, with all the fields and all objects detected in the fields. Loading data into the Science Archive could take a long time if the data were not clustered properly. Efficiency is important, since about 20 GB will be arriving daily.

The incoming data are organized by how the observations were taken. In the Science Archive they will be inserted into the hierarchy of containers as defined by the multi-dimensional spatial index (Figure 3), according to their colors and positions.

Data loading might bottleneck on creating the clustering units—databases and containers—that hold the objects. Our load design minimizes disk accesses, touching each clustering unit at most once during a load. The chunk data is first examined to construct an index. This determines where each object will be located and creates a list of databases and containers that are needed. Then data is inserted into the containers in a single pass over the data objects.

### Scalable Server Architectures

Accessing large data sets is primarily I/O limited. Even with the best indexing schemes, some queries must scan the entire data set. Acceptable I/O performance can be achieved with expensive, ultra-fast storage systems, or with many of commodity servers operating in parallel. We are exploring the use of commodity servers and storage to allow inexpensive interactive data analysis. We are still exploring what constitutes a balanced system design: the appropriate ratio between processor, memory, network bandwidth, and disk bandwidth.

Using the multi-dimensional indexing techniques described in the previous section, many queries will be able to select exactly the data they need after doing an index lookup. Such simple queries will just pipeline the data and images off of disk as quickly as the network can transport it to the astronomer's system for analysis or visualization.

When the queries are more complex, it will be necessary to scan the entire dataset or to repartition it for categorization, clustering, and cross comparisons. Experience will teach us the necessary ratio between processor power, main memory size, IO bandwidth, and system-area-network bandwidth.

Our simplest approach is to run a **scan machine** that continuously scans the dataset evaluating user-supplied predicates on each object [Acharya95]. We are building an array of 20 nodes. Each node has 4 Intel Xeon 450 Mhz processors, 256MB of RAM, and 12x18GB disks (4TB of storage in all). Experiments by Andrew Hartman of Dell Computers show that one node is capable of reading data at 150 MBps while using almost no processor time [Hartman98]. If the data is spread among the 20 nodes, they can scan the data at an aggregate rate of 3 GBps. This half-million dollar system could scan the complete (year 2004) SDSS catalog every 2 minutes. By then these machines should be 10x faster. This should give near-interactive response to most complex queries that involve single-object predicates.

Many queries involve comparing, classifying or clustering objects. We expect to provide a second class of machine, called a **hash machine** that performs comparisons within data clusters. Hash machines redistribute a subset of the data among all the nodes of the cluster. Then each node processes each hash bucket at that node. This parallel-clustering approach has worked extremely well for relational databases in joining and aggregating data. We believe it will work equally well for scientific spatial data.

The hash phase scans the entire dataset, selects a subset of the objects based on some predicate, and "hashes" each object to the appropriate buckets – a single object may go to several buckets (to allow objects near the edges of a region to go to all the neighboring regions as well). In a second phase all the objects in a bucket are compared to one another. The output is a stream of objects with corresponding attributes.

These operations are analogous to relational hash-join, hence the name [DeWitt92]. Like hash joins, the hash machine can be highly parallel, processing the entire database in a few minutes. The application of the hash-machine to tasks like finding gravitational lenses or clustering by spectral type or by redshift-distance vector should be obvious: each bucket represents a neighborhood in these high-dimensional spaces. We envision a non-procedural programming interface to define the bucket partition function and to define the bucket analysis function.

The hash machine is a simple form of the more general data-flow programming model in which data flows from storage through various processing steps. Each step is amenable to partition parallelism. The underlying system manages the creation and processing of the flows. This programming style has evolved both in the database community [DeWitt92, Graefe93, Barclay 95] and in the scientific programming community with PVM and MPI [Gropp98]. This has evolved to a general programming model as typified by a river system [Arpaci-Dusseau 99].

We propose to let astronomers construct dataflow graphs where the nodes consume one or more data streams, filter and combine the data, and then produce one or more result streams. The outputs of these rivers either go back to the database or to visualization programs. These dataflow graphs will be executed on a **river-machine** similar to the scan and hash machine. The simplest river systems are sorting networks. Current systems have demonstrated that they can sort at about 100 MBps using commodity hardware and 5 GBps if using thousands of nodes and disks [Sort].

With time, each astronomy department will be able to afford local copies of these machines and the databases, but for now, they will be a network service. The scan machine will be interactively scheduled: when an astronomer has a query, it is added to the query mix immediately. All data that qualifies is sent back to the astronomer, and the query completes within the scan time. The hash and river machines will be batch scheduled.

### Desktop Data Analysis

Most astronomers will not be interested in all of the hundreds of attributes of each object. Indeed, most will be interested in only 10% of the entire dataset – but different communities and individuals will be interested in a different 10%.

We plan to isolate the 10 most popular attributes (3 Cartesian positions on the sky, 5 colors, 1 size, 1 classification parameter) into small 'tag' objects, which point to the rest of the attributes. Then we will build a spatial index on these attributes. These will occupy much less space, thus can be searched more than 10 times faster, if no other attributes are involved in the query.

Large disks are available today, and within a few years 100GB disks will be common. This means that all astronomers can have a vertical partition of the 10% of the SDSS on their desktops. This will be convenient for targeted searches and for developing algorithms. But, full searchers will still be much faster on the server machines because the servers will have much more IO bandwidth and processing power.

Vertical partitioning can also be applied by the scan, hash, and river machines to reduce data movement and to allow faster scans of popular subsets.

We also plan to offer a 1% sample (about 10 GB) of the whole database that can be used to quickly test and debug programs. Combining partitioning and sampling converts a 2 TB data set into 2 gigabytes, which can fit comfortably on desktop workstations for program development.

### Distributed Analysis Environment

It is obvious, that with multi-terabyte databases, not even the intermediate data sets can be stored locally. The only way this data can be analyzed is for the analysis software to directly communicate with the Data Warehouse, implemented on a server cluster, as discussed above. Such an Analysis Engine can then process the bulk of the raw data extracted from the archive, and the user needs only to receive a drastically reduced result set.

Given all these efforts to make the server parallel and distributed, it would be inefficient to ignore IO or network bottlenecks at the analysis level. Thus it is obvious that we need to think of the analysis engine as part of the distributed, scalable computing environment, closely integrated with the database server itself. Even the division of functions between the server and the analysis engine will become fuzzy — the analysis is just part of the river-flow described earlier. The pool of available CPU's will be allocated to each task.

The analysis software itself must be able to run in parallel. Since it is expected that scientists with relatively little experience in distributed and parallel programming will work in this environment, we need to create a carefully crafted application development environment, to aid the construction of customized analysis engines. Data extraction needs to be considered also carefully. If our server is distributed and the analysis is on a distributed system, the extracted data should also go directly from one of the servers to one of the many Analysis Engines. Such an approach will also distribute the network load better.

### Summary

Astronomy is about to be revolutionized by having a detailed atlas of the sky available to all astronomers. If the SDSS is successful, it will be easy for astronomers to pose complex queries to the catalog and get answers within seconds, and within minutes if the query requires a complete search of the database.

The SDSS datasets pose interesting challenges for automatically placing and managing the data, for executing complex queries against a high-dimensional data space, and for supporting complex user-defined distance and classification metrics.

The efficiency of the instruments and detectors used in the observations is approaching 80%. The factor limiting resolution is the Earth atmosphere. There is not a large margin for a further dramatic improvement in terrestrial instruments.

On the other hand, the SDSS project is "riding Moore's law": the data set we started to collect today – at a linear rate – will be much more manageable tomorrow, with the exponential growth of CPU speed and storage capacity. The scalable archive design presented here will be able to adapt to such changes.

## *Acknowledgements*

We would like to acknowledge support from the Astrophysical Research Consortium, the HSF, NASA and Intel's Technology for Education 2000 program, in particular George Bourianoff (Intel).

## *References*

[Arpaci-Dusseau98] R. Arpaci-Dusseau, A. Arpaci-Dusseau, D. E. Culler, J. M. Hellerstein, D. A. Patterson. "The Architectural Costs of Streaming I/O: A Comparison of Workstations, Clusters, and SMPs", Proc. Fourth International Symposium On High-Performance Computer Architecture (HPCA), Feb 1998.

[Arpaci-Dusseau 99] R. H. Arpaci-Dusseau, E. Anderson, N. Treuhaft, D.E. Culler, J. M. Hellerstein, D. A. Patterson, K.Yelick, "Cluster I/O with River: Making the Fast Case Common." To appear in IOPADS '99. http://now.cs.berkeley.edu/River/

[Acharya95] S. Acharya, R. Alonso, M. J. Franklin, S. B. Zdonik, "Broadcast Disks: Data Management for Asymmetric Communications Environments." SIGMOD Conference 1995: 199-210.

[Barclay94] T. Barclay, R. Barnes, J. Gray, P. Sundaresan, "Loading Databases Using Dataflow Parallelism.", SIGMOD Record 23(4): 72-83 (1994)

[DeWitt92] D,.J. DeWitt, J. Gray, "Parallel Database Systems: The Future of High Performance Database Systems." CACM 35(6): 85-98 (1992).

[Csabai97] Csabai,I., Szalay,A.S. and Brunner,R.: Multidimensional Index For Highly Clustered Data With Large Density Contrasts, in Statistical Challenges in Astronomy II, eds. E. Feigelson and A. Babu, (Wiley), 447 (1997).

[GSC2] Greene, G. et al, " The GSC-I and GCS-II Databases: An Object Oriented Approach", in *New Horizons from Multi-Wavelength Sky Surveys*, eds B. McLean et al , (Kluwer), p.474 (1997)

[Gropp98] W. Gropp, S. Huss-Lederman, *MPI the Complete Reference: The MPI-2 Extensions, Vol. 2,* MIT Press, 1998, ISBN: 0262571234

[Graefe] G.Graefe, "Query Evaluation Techniques for Large Databases". ACM Computing Surveys 25(2): 73-170 (1993)

[Hartman98] Andrew Hartman of Dell Computer, private communication. Intel Corporation has generously provided the SDSS effort at Johns Hopkins with Dell Computers.

[Samet 1990a] H. Samet, *Applications of Spatial Data Structures: Computer Graphics, Image Processing, and GIS* , Addison-Wesley, Reading, MA, 1990. ISBN 0-201-50300-0.

[Samet 1990b] H. Samet, *The Design and Analysis of Spatial Data Structures*, Addison-Wesley, Reading, MA, 1990. ISBN 0-201-50255-0.

[Sort] The Sort Benchmark website: http://research.microsoft.com/barc/SortBenchmark/

[Szalay97] Szalay,A.S. and Brunner,R.J.: "Exploring Terabyte Archives in Astronomy", in New Horizons from Multi-Wavelength Sky Surveys, IAU Symposium 179, eds. B.McLean and D.Golombek, p.455. (1997).

[Szalay99] Szalay,A.S., Kunszt,P. and Brunner,R.A.: Hierarchical Sky Partitioning, Astronomical Journal, to be submitted, 1999.

[Wells81] Wells, D. C., Greisen, E. W., and Harten, R. H., FITS: A Flexible Image Transport System, Astronomy and Astrophysics Supplement Series, 44, 363-370, 1981.